\input{aipcheck}

\documentclass[
    ,final            
  ]
  {aipproc}

\layoutstyle{8x11double}

\begin{document}

\title{Is 44~Tau in the post-MS contraction phase?}

\classification{97.10.Sj}

\keywords      {
	 stars: variables: $\delta$ Sct --
	 stars: oscillations --
	 stars: individual: 44~Tau
}

\author{Patrick Lenz}{
  address={Institute of Astronomy, University of Vienna, T\"urkenschanzstr. 17, 1180 Vienna, Austria}
}

\author{Alexey A. Pamyatnykh}{
  address={Institute of Astronomy, University of Vienna, T\"urkenschanzstr. 17, 1180 Vienna, Austria} 
  ,altaddress={Copernicus Astronomical Center, Polish Acad. Sci., Bartycka 18, 00-716 Warsaw, Poland} 
  ,altaddress={Institute of Astronomy, Russian Acad. Sci., Pyatnitskaya Str. 48, 109017 Moscow, Russia} 
}

\author{Michel Breger}{
  address={Institute of Astronomy, University of Vienna, T\"urkenschanzstr. 17, 1180 Vienna, Austria}
}

\begin{abstract}
The evolutionary stage of the $\delta$~Scuti star 44~Tau has been unclear. Recent pulsation studies have claimed both main sequence and post-main sequence expansion models. A new photometric study increased the number of detected frequencies in 44~Tau to 49, of which 15 are independent modes. We now find that a previously ignored third possibility, the post-main sequence contraction phase, is in excellent agreement with the observed frequency range, as well as the frequency values of all individual radial and nonradial modes. These results resolve the previous disagreements in the literature and exemplify that asteroseismology can determine the evolutionary status of a star.
\end{abstract}

\maketitle


\section{Introduction}

44~Tau (spectral type: F2~IV) is special amongst the $\delta$~Scuti stars because it rotates with an exceptional low equatorial rotation rate of $V_{\rm rot}$~=~3~$\pm$~2~km~s$^{-1}$ (Zima et al. 2007 \cite{zima2007}). The measured $\log g$ value of 3.6 $\pm$ 0.1 does not allow for an unambiguous determination of the evolutionary status of 44~Tau. 

Recently, Breger \& Lenz (2008) \cite{bregerlenz2008} analyzed additional photometry of 44~Tau and found two new independent frequencies at 7.79 and 5.30 cd$^{-1}$. The latter frequency extends the range of observed frequencies to lower values. As stated by Lenz et al. (2008) \cite{lenz2008}, the observed frequency ranges can be used to decide between main sequence (hereafter MS) and post-MS models. Therefore, this new frequency provides an important observational constraint to our asteroseismic models. We reexamined pulsation models for all possible evolutionary stages taking into account all 15 independent frequencies and, if available, their mode identifications.

\section{Asteroseismic modelling}

Asteroseismic models of 44~Tau on the MS and in the expansion phase of the post-MS evolution were already examined in an earlier paper by Lenz et al. (2008) \cite{lenz2008}. This paper also describes the evolutionary and pulsation codes used in this work. Prior to this study a MS model of 44~Tau was computed by Garrido et al. (2007) \cite{garrido2007}. These papers also provide mode identifications. For the ten dominant modes the spherical degree is known. For eight modes the azimuthal order could be determined in a spectroscopic study by Zima et al. (2007) \cite{zima2007}. Moreover, these authors measured the abundances of photospheric elements in 44~Tau and found that they are close to the solar values.

Fortunately, the mode identifications put tight constraints on asteroseismic models. Two radial modes, four dipole modes and four quadrupole modes are observed in 44~Tau. From the frequencies of the radial modes the mean density and 
the mass can be inferred by means of Petersen diagrams (Petersen \& J{\o}rgensen 1972 \cite{petersen1972}). Two of the $\ell$~=~1 modes are much closer than the expected frequency separation of acoustic modes which indicates that we observe an avoided crossing (Aizenman et al. 1977 \cite{aizenman1977}). Such an observation puts strong constraints on the extent of overshooting from the convective core as shown by Dziembowski \& Pamyatnykh (1991) \cite{wad1991}. 

\begin{figure}
  \includegraphics[height=.9\textheight]{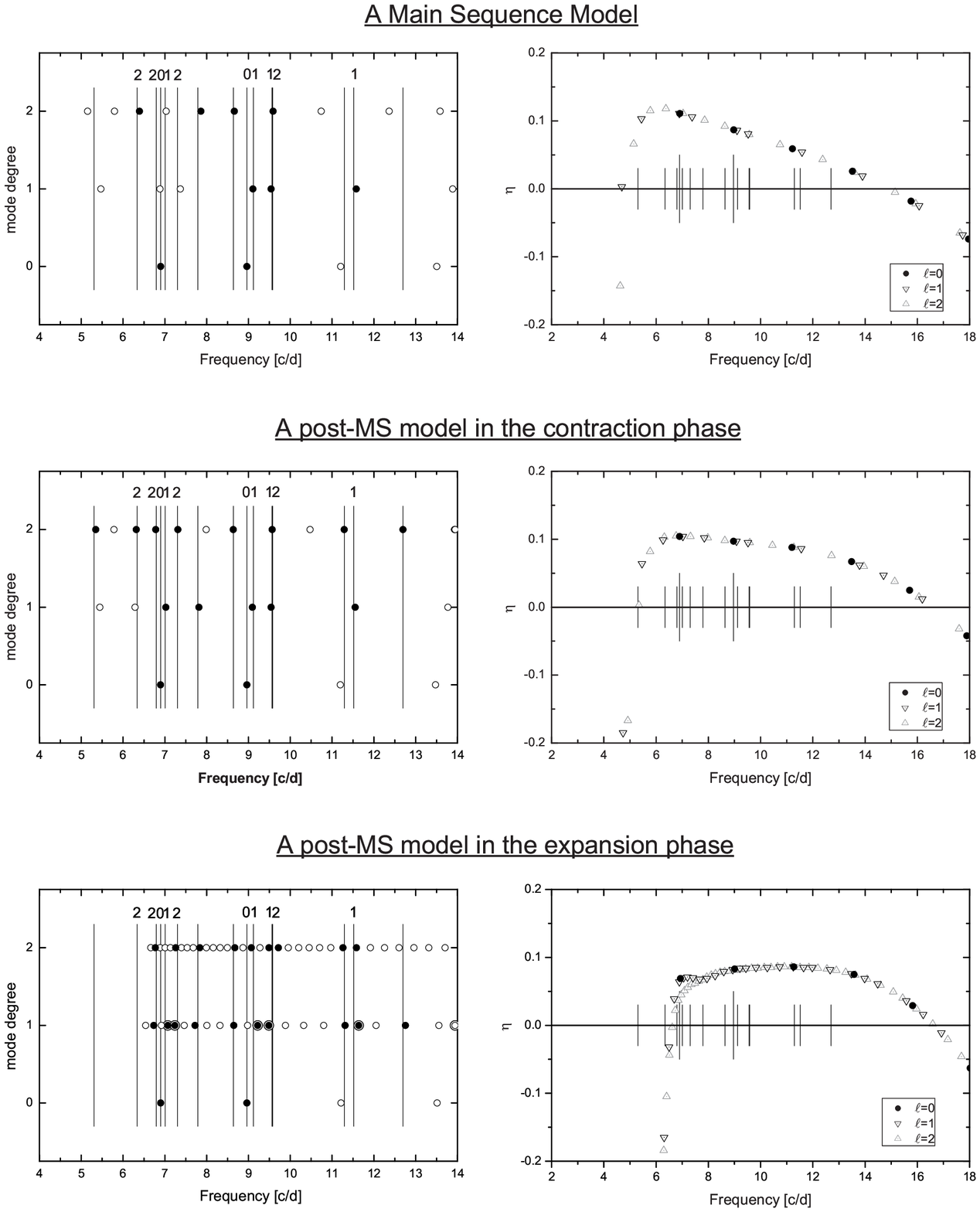}
  \caption{The diagrams on the left compare observed frequencies (vertical lines) and predicted unstable frequencies (circles) for models in different evolutionary stages. If predicted frequencies match observed frequencies they are marked as filled circles. The numbers above the vertical lines indicate the measured spherical degrees of the modes. For the model in the post-MS expansion stage of evolution, partially trapped $\ell$~=~1 modes are indicated with additional concentric black circles. The rotational splitting is about the same size as the symbols. The diagrams on the right show the corresponding instability parameter, $\eta$, vs. frequency. Positive values of $\eta$ indicate unstable modes. The vertical lines mark the position of observed frequencies (with longer lines for the two observed radial modes).}
  \label{fig:main}
\end{figure}

We tried to find the best fit of all observed frequencies by examining pulsation models in all possible evolutionary stages. We also investigated a previously ignored possibility: a model in the overall contraction phase at the end of hydrogen burning in the core.
The pulsation models were computed using OPAL opacities (Iglesias \& Rogers 1996 \cite{iglesias1996}) and the GN93  element mixture (Grevesse \& Noels 1993 \cite{gn93}). Our nonadiabatic pulsation code also predicts whether a mode is unstable or not. The theoretical frequency spectra of the best models in different evolutionary stages are shown in Fig.~\ref{fig:main}. 

MS models generally exhibit a poor agreement between the observed $\ell$~=~2 modes at 6.34, 6.80 and 7.30~cd$^{-1}$ and the corresponding frequencies of the theoretical modes. This problem could not be solved by changing the chemical composition and other input parameters within the limits set by the measurements. Moreover, the predicted luminosity and the effective temperature of this model are lower than the observed values (see Lenz et al. 2008 for more detailed parameters of this model). However, the predicted instability range is in excellent agreement with observations.

Post-MS models in the expansion phase do not predict instability for the two observed modes at lowest frequencies. Moreover, many more frequencies are predicted than observed. This can be partly explained by a mode selection mechanism such as partial trapping of modes in the stellar envelope. However, mode trapping is not effective for quadrupole modes and, hence, does not explain the observed frequencies of the $\ell$~=~2 modes satisfactorily.

Finally, we also examined pulsation models in the contraction phase after the TAMS. Towards the end of the MS evolution hydrogen core burning gradually becomes less efficient because of the decreasing hydrogen abundance in the core. The star reacts with an overall contraction which increases its central temperature. At some point shell hydrogen burning becomes the dominant source of nuclear energy which causes the outer layers to expand again.

\begin{figure}
  \includegraphics[height=.25\textheight, bb=65 20 240 225]{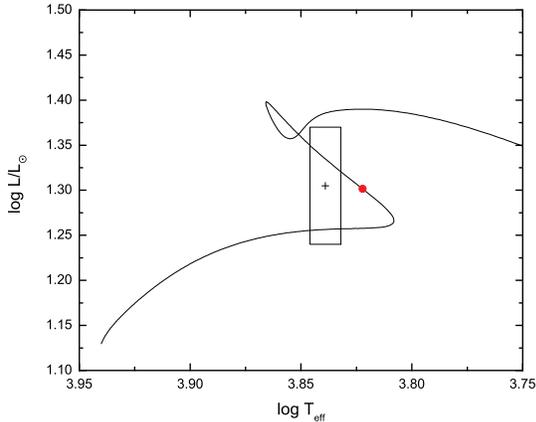}
  \caption{Evolutionary track and position of the pulsation model with the excellent frequency fit in the post-MS contraction phase in the HR diagram.}
  \label{fig:hrd}
\end{figure}

  The time spent in this contraction phase is approximately 10 times shorter than on the main sequence for the same range of effective temperatures. However, both the excellent fit of predicted and observed frequencies and the predicted range of unstable modes indicate that 44~Tau is in the post-MS contraction phase.
The position of this model in the HR diagram is shown in Fig.~\ref{fig:hrd}. The  predicted effective temperature is cooler than the value derived from photometry. However, the standard deviation of the photometric effective temperature shown in the diagram is 100K which may be optimistic. The predicted luminosity is in good agreement with the values derived from Hipparcos. 

The parameters of this model are: $T_{\rm eff}$~=~6640~K, $\log L$~=~1.302, $\log g$~=~3.682 and the mass is 2.021 M$_{\odot}$.
The pulsation model was computed with a hydrogen mass fraction of X~=~0.75. The higher hydrogen abundance helps to minimize the frequency separation between the dipole modes that undergo an avoided crossing and, therefore, provides a better fit than models with X~=~0.70. The metal mass fraction was assumed to be Z~=~0.02, the mixing length parameter, $\alpha_{MLT}$~=~0.2, and the overshooting parameter, $\alpha_{\rm ov}$~=~0.195. 
The model parameters of the MS and post-MS model in the expansion phase are the same as those given in Lenz et. al (2008).

\section{Conclusions}

We reexamined pulsation models of 44~Tau taking into account two new independent frequencies found by Breger \& Lenz (2008) \cite{bregerlenz2008} and also considered the hitherto neglected possibility that 44~Tau is in the contraction phase at the end of hydrogen-core burning. By comparing the predicted frequency spectra in all three possible evolutionary stages we find that for a post-MS contraction model all 15 observed frequencies are well reproduced,
 the total number of predicted modes is in good agreement with observations, and
 the predicted instability of the modes matches the observed range of frequencies.
The differences between models in different evolutionary stages are mainly due to different conditions in the stellar interior which can only be inferred by asteroseismology. In the case of 44~Tau, asteroseismology successfully retrieved the evolutionary status.


\begin{theacknowledgments}
This investigation has been supported the Austrian Fonds zur F\"orderung der wissenschaftlichen Forschung (project P21830).
 AAP acknowledges partial financial support from the Polish MNiSW grant No. N N203 379636.
\end{theacknowledgments}



\bibliographystyle{aipproc}   


\begin{thebibliography}{9}

\bibitem{aizenman1977} 
        M. Aizenman, and P. Smeyers, A. Weigert, \emph{A\&A}, \textbf{58}, 41 (1977)

\bibitem{bregerlenz2008}
        M. Breger, and P. Lenz, \emph{A\&A} \textbf{488}, 643 (2008)

\bibitem{wad1991}
        W.~A. Dziembowski, A.~A. Pamyatnykh, \emph{A\&A} \textbf{248}, 11 (1991)

\bibitem{garrido2007}
        R. Garrido, and J.~C. Su\'{a}rez, and A. Grigahc\`{e}ne et al., \emph{CoAst} \textbf{150}, 77 (2007)

\bibitem{gn93}
	N. Grevesse, and  A. Noels,
        in \emph{Origin and Evolution of the Elements}, edited by Prantzos N., Vangioni-Flam E. and Casse M., CUP, 1993, p. 15


\bibitem{iglesias1996} 
    C. A. Iglesias, and  F. J. Rogers, \emph{A\&A} \textbf{464}, 943 (1996) 

\bibitem{lenz2008}
        P. Lenz, and A.~A. Pamyatnykh, and M. Breger, and V. Antoci, \emph{A\&A} \textbf{478}, 855 (2008)

\bibitem{petersen1972}
	J.~O. Petersen, and H.~E. J{\o}rgensen, \emph{A\&A} \textbf{17}, 367 (1972)

\bibitem{zima2007}
        W. Zima, and H. Lehmann, and Ch. St\"utz, and I.~V. Ilyin, and M. Breger, \emph{A\&A} \textbf{471}, 237 (2007)

\end{thebibliography}


\end{document}